# Opening a new window to other worlds
# with spectropolarimetry

**Maren Mohler · Johannes Bühl · Stephen Doherty · Siegfried Eggl ·
Vera Theresa Eybl · François Farago · Aleksandar Jaćimović ·
Lars Hunger · Nynne L. B. Lauritsen · David Ludena ·
Martina Meisnar · Alexander Reissner · Nicolas Sarda ·
Benjamin Toullec · Meritxell Viñas Tió**

**Abstract** A high level of diversity has already been observed among the
planets of our own Solar System. As such, one expects extrasolar planets to
present a wide range of distinctive features, therefore the characterisation
of Earth- and super Earth-like planets is becoming of key importance in

M. Mohler (✉)
Institute for Astrophysics, University of Göttingen, Göttingen, Germany
e-mail: marenmohler@gmx.de

J. Bühl
Institute of Applied Optics, Friedrich-Schiller-University Jena, Jena, Germany

S. Doherty
Dept. of Experimental Physics, National University of Ireland (NUI)—Maynooth,
Maynooth Co., Kildare, Ireland

S. Eggl · V. T. Eybl
Institute for Astronomy, University of Vienna, Vienna, Austria

F. Farago
ASD, IMCCE-CNRS UMR8028, Observatoire de Paris, Paris, France

A. Jaćimović
Utrecht University, Utrecht, The Netherlands

L. Hunger
Institut für Astro- und Teilchenphysik, Universität Innsbruck, Innsbruck, Austria

N. L. B. Lauritsen
Niels Bohr Institute, University of Copenhagen, Copenhagen, Denmark

D. Ludena
Deutsches Zentrum für Luft- und Raumfahrt (DLR), Institut für Planetenforschung,
Berlin, Germany

scientific research. The SEARCH (**S**pectropolarimetric **E**xoplanet **A**tmosphe**R**e **CH**aracerisation) mission proposal of this paper represents one possible approach to realising these objectives. The mission goals of SEARCH include the detailed characterisation of a wide variety of exoplanets, ranging from terrestrial planets to gas giants. More specifically, SEARCH will determine atmospheric properties such as cloud coverage, surface pressure and atmospheric composition, and may also be capable of identifying basic surface features. To resolve a planet with a semi major axis of down to 1.4 AU and 30 pc distant SEARCH will have a mirror system consisting of two segments, with elliptical rim, cut out of a parabolic mirror. This will yield an effective diameter of 9 m along one axis. A phase mask coronagraph along with an integral spectrograph will be used to overcome the contrast ratio of star to planet light. Such a mission would provide invaluable data on the diversity present in extrasolar planetary systems and much more could be learned from the similarities and differences compared to our own Solar System. This would allow our theories of planetary formation, atmospheric accretion and evolution to be tested, and our understanding of regions such as the outer limit of the Habitable Zone to be further improved.

# 1 Introduction

In recent years, efforts to find extrasolar planets steadily increased and continue to do so. Advanced telescope technologies, such as those in Chile, allowed us to find a variety of exoplanets. The HARPS spectrograph at the 3.6-m telescope at La Silla is one of the leading ground-based data sources of exoplanet radial velocity curves and has already detected and confirmed a significant number of exoplanet targets [57, 59]. The launch of the CoRoT satellite in 2006 offered new possibilities for detecting exoplanets using high resolution photometry to measure transit curves. Together with ground-based

M. Meisnar
Institute for Solid State Physics, Vienna University of Technology, Vienna, Austria

A. Reissner
AIT—Austrian Institute of Technology, Wien, Austria

N. Sarda
Astrium Ltd, Hertfordshire, UK

B. Toullec
Supaero—Institut Supérieur de l'Aéronautique et de l'Espace, Toulouse, France

M. Viñas Tió
Escola Politécnica Superior de Castelldefels (EPSC), Barcelona, Spain

follow-up observations with HARPS the detection of the rocky Super-Earth planet CoRoT-7b was recently confirmed [46, 61]. High expectations have been raised by NASAs Kepler spacecraft, launched in March 2009, which will observe over 100,000 stars in the Cygnus constellation [12] with the aim of identifying specifically Earth and Super-Earth-sized planets.

Several more projects are in the pipeline. One example being ESPRESSO (Echelle Spectrograph for PREcision Super Stable Observations), a new generation instrument for the European Southern Observatory's VLT. It will combine the high stability of HARPS and the efficiency of UVES at the VLT and is planned to be the precursor of CODEX, a high resolution spectrograph for the European ELT [48]. A space mission with the primary objective of finding Earth-sized planets outside our solar system is NASAs SIM (Space Interferometry Mission) [78]. A ground-based version of this instrument, PRIMA (Phase Referenced Imaging and Microarcsecond Astrometry), is installed at the VLTI (the Very Large Telescope Interferometer) [15]. These two missions are expected to identify several more Earth-sized exoplanets via interferometry. Furthermore, ESA's GAIA mission scheduled for 2011 will provide a huge catalogue of approximately 1 billion stars up to magnitude 20 and will include astrometric, spectroscopic and radial velocity data. As such, one may assume that many new targets for further exoplanetary research will be available in a few years [44, 51]. Another two concepts being considered in Europe and Japan are the PLATO and SPICA projects.

The next logical step following detection is the characterisation and analysis of the catalogued exoplanets. A detailed analysis of the connection between planet features and host star characteristics will also be made possible by future detections in conjunction with the GAIA data. The characterisation of a large number of varied targets will be invaluable for the further development of present models. This characterisation of exoplanets in terms of their atmospheric properties is the purpose of the mission study SEARCH presented in this paper. The mission concept was developed during the Summer School held in Alpbach during July 21–30, 2009 with the title: Exoplanets: Discovering and Characterising Earth Type Planets and further studied during a Post Alpbach week at the Space Research Institute of the Austrian Academy of Sciences during November 24–27, 2009, in Graz, Austria.

Spectropolarimetry is still a relatively new technique, however the potential of the technique is being realised. This is reflected in similar approaches being suggested for both ground- and space-based projects [5, 67], which are currently on the exoplanetary communities' drawing-boards.

## 2 Science case

### 2.1 The method: spectropolarimetry

Polarimetry is a powerful technique for enhancing the contrast between a star and an exoplanet. Indeed, the light from an inactive star integrated over its whole disk is usually unpolarised. However, when scattered by a planetary atmosphere and/or surface, the light becomes polarised, and thus the reflected light which reaches us presents a significant degree of polarisation. Therefore, measuring the polarisation degree of the incoming light very precisely, is equivalent to gaining five orders of magnitude in contrast [37]. Additionally, the dependence of the degree of polarisation on the wavelength presents characteristic features which can be used to infer the structure of the scattering atmosphere/surface. Significant work has been carried out to produce characteristic spectra of a wide range of planetary atmospheres and surfaces [27, 64, 69, 72–75].

One of the characteristics of planetary atmospheres which could be easily obtained through spectropolarimetry is the surface pressure. The Rayleigh scattering cross-section is $\propto \lambda^{-4}$. For low surface pressures, the atmospheric density allows only for single scattering events at short wavelengths, and the wavelength dependence of the Rayleigh cross-section suppresses the scattering of longer wavelengths. The degree of polarisation thus decreases with wavelength. For higher surface pressures, the atmospheric density will allow multiple scattering events at short wavelengths, which destroy the polarisation in this domain, while longer wavelengths will undergo only single scattering events and are significantly polarised.

Numerical simulations of different kinds of planetary atmospheres and surfaces, together with the observation of the planets of the Solar System, have shown that many planetary features have clear signatures in the polarisation spectrum: cloud coverage, cloud particle size and shape, ocean coverage, and other surface coverage properties such as the existence of potential vegetation. By comparing such simulations with measurements one is able to characterise these properties and acquire a quite detailed picture of the observed atmosphere.

The polarisation spectrum has broad features which do not require a high spectroscopic resolution: $R \approx 70$–80 is sufficient to resolve them [32]. As such, a rough characterisation of an observed atmosphere is possible after only a short amount of time. The integration times needed to achieve a sufficient signal-to-noise ratio at this low resolution would allow for the rapid exploration of about 50 targets with different phase angles for each target, as will be described below. The variation of the polarisation signal with respect to phase angle provides information on the orbital inclination of the target exoplanet, which is a missing parameter in radial-velocity data. In turn, the orbital inclination allows for a more precise determination of the planetary masses (see Fig. 1, right). After a first orbital sampling phase at low spectral resolution ($R \approx 70$) high-interest exoplanets can be identified, on which further analysis could be carried out with longer exposure times and a higher spectral resolution. This would yield data on the the absorption lines in the full flux spectrum and on the chemical composition of the atmosphere.

The power of spectropolarimetry comes from its utilisation of both the spectral information and also polarised nature of the signal, which is usually lost when only a flux spectrum is taken. Limiting measurements to the flux spectrum would provide data only on the chemical composition of the atmosphere. It would be difficult to infer many important characteristics of the observed atmosphere, for example cloud properties and the surface pressure. That means it would be hard to distinguish between a thick Venus-like atmosphere and a thin Mars-like atmosphere. Basic surface features would also be inaccessible. Since this project aims at characterising planetary atmospheres, the aforementioned parameters are also of interest. For this reason spectropolarimetry was chosen as the preferred method.

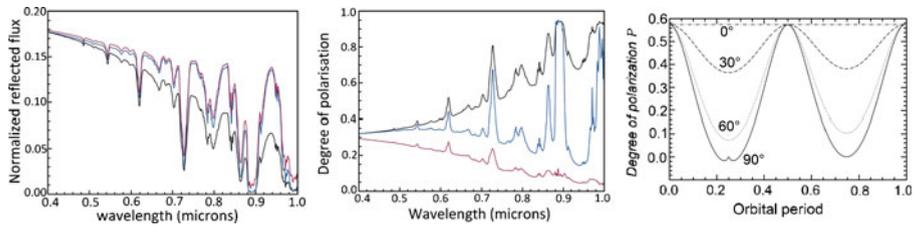

**Fig. 1** The flux (*left*) and degree of polarisation (*middle*) of starlight reflected by three Jupiter-like exoplanets for the phase angle alpha 90. Planetary model atmosphere 1 (*black line*) contains only molecules, model 2 (*blue line*) is similar to model 1 except for a tropospheric cloud layer and model 3 (*red line*) is similar to model 2 except for a stratospheric haze layer. The flux spectra exhibit very similar features which are hard to interpret, while the polarisation spectra have distinctive slopes, giving easy access to atmospheric structure. Found in Stam et al. [75]. *Right* Variation of the polarisation degree with respect to planetary phase angle depending on orbital inclination for a Jovian planet [75]

## 2.2 Understanding the diversity of Exoplanets

Spectropolarimetry makes it possible to observe the distinct characteristics of planets around stars other than the Sun for the first time. Investigating the atmospheric and surface properties of a sample of extrasolar planets around different types of stars allows one to test current models of atmospheric evolution and habitability concepts. The proposed mission would be able to resolve a planet as close to its host star as 0.5 AU in a distance of 10 pc, and as close as 1.4 AU in a distance of 30 pc. That provides a wide range of planets from terrestrial Earth-like planets to gas giants, which would be sufficient for statistical analysis.

### 2.2.1 Exploring the diversity of rocky and icy worlds

For planets in the Earth to super-Earth mass range a huge variety of possible compositions can be expected. Planets unknown in the Solar System have been discovered, providing a possibility to check our understanding of planetary formation and evolution. The suggested types of low-mass planets range from Earths and super-Earths to ocean worlds, icy planets and even rocky cores of Neptune-type bodies. Models predict an upper radius limit for terrestrial-type planets with a given mass [3, 79], separating dry rocky compositions from planets containing 10% and more water and other volatiles.

At the moment there is little knowledge of the early stages of planetary evolution. Numerical simulations have been performed, showing that planets may form with a much higher water content than is observed in the Solar System [3, 62]. Furthermore, it has been suggested that a super-Earth with the same bulk composition as Earth should be completely covered in water, since the surface of the planet scales with the square of the radius, while the volume follows a cubic law [49]. This could make ocean planets very common.

Above the terrestrial-type mass-radius limit, water worlds cannot be distinguished from planets with a significant atmospheric layer of H/He based on planetary radius and mass measurements alone [1]. Planets within this regime could be either super-Earths or mini-Neptunes [4]. It's possible to draw conclusions about the bulk composition from the mass/radius relationship (Fig. 2), but there are uncertainties regarding helium and carbon planets [71]. To make conclusions about the applicability of planetary composition models, one has to gain additional information about the planets atmospheric and surface properties. Williams and Gaidos [82] suggest the direct detection of the oceans glint, which makes a partially water-covered planet appear brighter near crescent phase. A planet that is covered completely with water appears darker than a Lambertian disk. Planets formed beyond the ice line, migrating inwards, as well as planets with a high concentration of volatiles delivered from impacts of planetesimals, end up as icy worlds, like Titan in the Solar System. Carbon planets [39] and rocky Neptune-cores are examples of exoplanets yet unknown. Neptune-like bodies could lose all of their hydrogen atmospheres due to stellar wind and high XUV flux either in the early phases of stellar

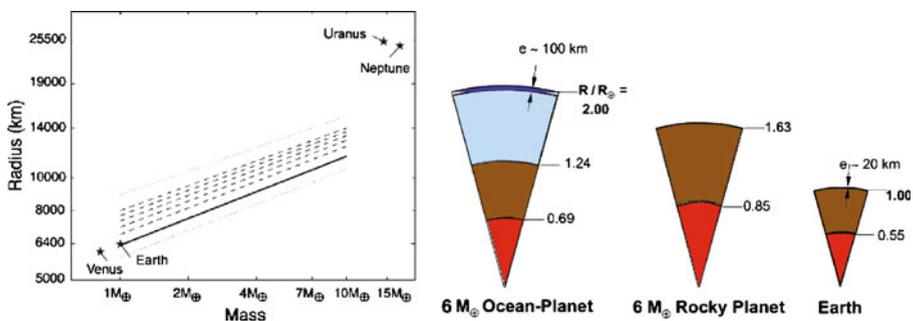

**Fig. 2** *Left* Mass-radius relationship for ocean and rocky massive planets. The *solid black line* is the power-law relationship for terrestrial planets with 1–10 $M_\oplus$. The *dashed lines* above progressively represent the relationship for planets with 10, 20, 30, 40, and 50% $H_2O$. This family of planets has a fixed mantle-to-core proportion of 2:1. The minimum and maximum planetary radius relations with mass are shown as *dotted lines*. Venus, Earth, Uranus, and Neptune are shown for reference [79]. *Right* From *left* to *right*: calculated internal structure of a 6 $M_\oplus$ ocean planet, a 6 $M_\oplus$ rocky planet and the Earth, respectively [45]

evolution or during inward migration. These rocky bodies could evolve to a new species of terrestrial-type planets [38, 43].

While there is no theory to date regarding the suitability of water and ice worlds for hosting life, models exist to quantify the habitability of terrestrial-type, rocky planets. The Habitable Zone [36] (Fig. 5, right) is the concept widely used to describe the region around a star where life could possibly exist. Our concept of habitability is strongly based on the conditions suitable for life as we know it on Earth. Although used in a broader sense recently, the Habitable Zone (HZ) is originally defined as the distance from a star, where liquid water on a planets surface can exist. It is calculated based on the stellar flux received by a planet with an Earth-like atmosphere.

The inner edge of the HZ is determined by evaporation of oceans caused by a runaway greenhouse effect and the subsequent loss of water to space. This is believed to have happened on Venus, making it a dry planet with a dense $CO_2$ atmosphere [35]. The original paper by Kasting gives a conservative estimate for the inner border of the habitable zone of a G-star of 0.95 AU. The outer edge however is not as clearly defined, as the influence of $CO_2$ clouds is not yet fully understood. Estimates for the outer border of the HZ for a G-type star range from 1.37 AU [36] to 2.4 AU [20, 54]. Besides liquid water, a second factor for habitability is the presence of an atmosphere dense enough to maintain a stable surface temperature by the greenhouse effect. The greenhouse effect is caused by atmospheric gases like $H_2O$, $CO_2$ and $CH_4$, which are very efficient absorbing in the infrared, but not in the visible range of the spectrum. On Earth, the amount of $CO_2$ in the atmosphere is controlled by the carbonate-silicate-cycle, acting as a thermostat. Volcanic activity releases $CO_2$ into the atmosphere, which is washed out by weathering and subsequently gets buried in ocean sediments. At the subduction zones, these are transferred into the mantle. Therefore, plate tectonics is considered to be a major factor in making a planet a potential habitat, since it provides an efficient mechanism enabling a $CO_2$ cycle [26, 42].

At very early stages of planetary evolution chemical reactions in the freshly formed crust of the planet produce methane and ammonia, creating the planet's atmosphere together with water vapour. Since young stars are most active, photolysis in the upper layers of the atmosphere breaks up the molecules of $CH_4$, $NH_3$ and $H_2O$. As well as outgassed $CO_2$, hydrogen is lost to space, while carbon and oxygen react to form $CO_2$. So all terrestrial-type planets should start out with a $CO_2$-rich atmosphere [40–42]. If plate tectonics is active on a planet in the habitable zone, $CO_2$ from the atmosphere is recycled into the planets mantle, leaving behind an atmosphere dominated by nitrogen and depleted of carbon dioxide, similar to Earth [36]. Accordingly, the amount of $CO_2$ and nitrogen, respectively, in a planets atmosphere can be an indicator of the presence of plate tectonics.

There has been an ongoing discussion about whether Super-Earths are to be expected to maintain plate tectonics, and if there is a mass limit on plate tectonics. Some models predict stronger convection on terrestrial planets larger than Earth [80]. Others state that the increased mantle depth reduces convection thus decreasing the likelihood of plate tectonics [58]. A sufficiently large sample of terrestrial planets should allow more stringent conclusions. Since nitrogen can not be observed in the spectral lines, the observation of non-$CO_2$ versus $CO_2$-rich atmospheres in different environments provides insights into the conditions for habitability.

Using spectropolarimetry, it is possible to not only determine the composition of the atmosphere, but also distinguish between dense and thin/no atmospheres. This information can be used to separate Earth-like, Mars-like and Venus-like atmospheres. Weather phenomena like clouds can be observed, which helps in gaining a deeper understanding of the conditions present at a planets surface.

Investigating samples of low-mass planets (1–30 $M_\oplus$) around different types of stars in various stages of evolution, the concept of habitability can be tested and refined, giving us the unique opportunity to set the Solar System in context. Observations of surface features like oceans and continents as well as atmospheric properties will provide the data needed to verify and improve currently adopted models.

*Features that can be studied*   As mentioned before the analysis of atmospheres is based on the analysis of different features in the obtained spectrum. Table 1 shows several atmospheric features in the visible region of the spectrum.

One of the important features mentioned in Table 1 is the so called "red edge". A closer look at the Earth's spectrum reveals a high jump in it around 700 nm, which is the result of the chlorophyll in plants on Earth. The plants green colour is due to the absorbtion of light between $\approx$450–680 nm by the chlorophyll. As Fig. 3 (right) shows, the plants stop absorbing the light around 700 nm and the relative reflectivity spikes in this wavelength region. The expectation to see this feature in spectra of exoplanets is based on the assumption that plant life on other planets is similar to that on Earth. Unfortunately, the probability of observing this particular feature with the proposed telescope remains low.

### 2.2.2 Gas giants

The diversity among extrasolar planets is not limited to that of rocky planets. Extrasolar giant planets (EGPs), the most abundant among the currently catalogued exoplanets, offer, like the rocky ones, a considerable degree of variety due to the differences in their mass, age, radius, distance from host star, the stellar type and luminosity and the properties of the planetary system. The characterisation of the known EGPs combined with the knowledge of the Solar System giants would definitely improve our understanding of how planets and planetary systems form and evolve. The ongoing and scheduled missions for exploring the Solar System's gas giants, Cassini for Saturn and Juno for Jupiter and EJSM for Galilean moons will provide important information on the composition and the properties of the outer atmospheres, which will provide improved models of the gas giants. On the other hand, the characterisation of EGPs would be useful for making valuable comparisons between giant planets, placing in that way the Solar System's giants in a more general context. EGPs are easier to observe than rocky planets, considering that the flux from a Jupiter sized planet is about 100 times larger than the flux from an Earth sized one and that the number of available EGP targets is also much higher than the number of lower mass ones.

**Table 1** Line position of atmospheric features in the visible range

| Feature | Wavelength (nm) |
|---|---|
| $H_2O$ | 514, 575, 610, 730, 830 |
| $O_2$ | 626, 688, 767 |
| $O_3$ | Chappuis bands between 375–650 |
| $NH_3$ | 552, 647 |
| $CH_4$ | 486, 543, 576, 595, 629, 681, 703, 727, 790, 840, 864, 889 |
| Red edge | $\approx$700 |

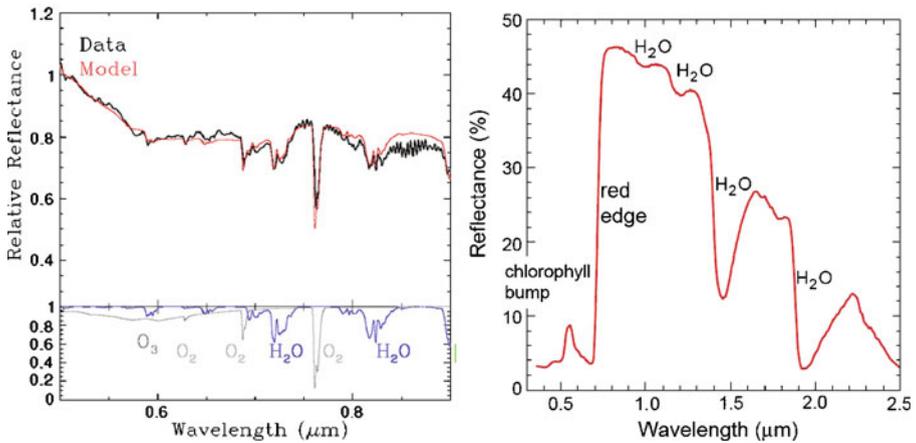

**Fig. 3** *Left* Observed reflectivity spectrum in the visible of the Earth, as determined from earthshine. The data is shown in *black* and the model in *red*. The reflectivity scale is arbitrary [32]. *Right* Reflectance of a deciduous leaf with visible red edge [70]

The EGPs, unlike their stars, have atmospheric temperatures that are low enough for the chemical composition to be of overriding importance. The atmospheric levels of EGPs that are emitting detectable radiation are the thin outer layers of molecules that control the absorption and emission spectra and the cooling rate. The dominant constituents are hydrogen and helium molecules, but depending on the above listed properties of the planet the presence of other molecules can be inferred or even directly observed using advanced spectroscopy. Spectropolarimetric observations of EGPs, i.e. the observations of the flux and the state of polarisation of planetary radiation from EGPs could allow going further in the planets characterisation than only estimating its size and distance from the star, which sum up the basic steps in making a first estimate of how comparable the system is to the Solar System. Information about the structure and composition can be inferred from the reflected visual and near infra red spectrum. The reflected and emitted spectra of EGPs are mainly products of the scattering of incident light from clear atmospheric gases above the clouds (Rayleigh scattering), and form the aerosols, cloud particles, and the absorption and emission of gaseous absorbers in the clouds and hazes (Mie scattering) [50]. Reflected starlight from the EGPs will, thus, in general be polarised in comparison to the overall integrated starlight of the stellar disc, and the degree of polarisation is, as in rocky planets, strongly dependent on the composition of the planetary atmosphere. As an example, in their spectropolarimetric studies, Joos et al. find that methane bands enhance polarisation in the red and near IR [28]. The photons that undergo multiple scatterings or penetrate deeply into the atmosphere to the cloud layers lead to strong methane absorptions which further reduce the unpolarised scattered light. Joos et al. also conclude that this enhanced polarisation can be detected in EGPs and they recommend the observation of the strong methane bands for all exoplanet characterisation. However, Stam et al. concluded that the degree of polarisation can as well decrease in the absorption bands for different models of EGPs atmospheres [75]. The degree of polarisation of reflected starlight from an EGP is strongly dependent on the ratio of single scattered to the multiply scattered light. With little or no absorption in the atmosphere, multiply scattered light is observed with lower degree of polarisation. With some absorption by gas or cloud particles, most of the observed light is single scattered and therefore has a high degree of polarisation. With a lot of absorption, single scattered light will originate at different altitudes in the atmosphere and hence is likely to come from different types of particles. The single scattered light can therefore, in some models of the atmosphere, have a lower degree of polarisation then that of the multiply scattered light [73]. Next to the methane, principal absorbers are water and ammonia and hydrogen dipoles, as given in the Fig. 4 for the solar giants. Further on, the degree of polarisation followed in continuum for increasing wavelengths contains the differential information on particles at various altitudes in the atmosphere, as higher wavelengths penetrate deeper in the atmosphere. The degree of polarisation is a relative measure, and therefore free of the conditions of the star-planet-observer system which allows the extraction of the information on EGPs atmospheres even though the other, system characterising variables are not correctly measured [75]. For the discussion on EGPs it is common to split them on the hot and cold Jupiters. Hot Jupiters are the giant plants that inhabit regions very close to their stars, and therefore have intrinsically different features and are beyond the scope of the mission. However there exists a significant step variation between the true representatives of cold and hot Jupiters. They are usually, depending on their T-P profile and distance from the star, divided by the Sudarsky exosolar planet classification into five different classes, three out of which Class I, II and III are still within the scope of this mission [76]. The noted classes comprise the giant planets that are at least 0.5 AU away from the star and have equilibrium temperatures on average below 500 K. These classes also depend on the type of the absorbers present in the atmosphere. For the colder Class I it would be methane or ammonia clouds, for a slightly warmer Class II it would be water vapour clouds, with sulphuric acid clouds in the more extreme cases. The Class III planets are expected not to form global cloud cover as they are too warm to form water vapour clouds and too cold for a silicate layer to reach the upper troposphere. Hence, they would appear as featureless blue globes due to Rayleigh scattering, which is why Sudarsky names them clear. The Class III EGPs should, therefore, like Class II EGPs show significant absorptions of water, methane and collision induced molecular hydrogen, but as well some alkali metals like Na and K could appear in the visible. As all these lines are detectable in red and near infrared region, they are well within the scope of this mission. Finally, in accordance with the core accretion theory, planets that are more massive than 10 $M_{\oplus}$ should have a gaseous envelope from their surrounding nebula [50]. During planetary evolution the bombardment of planetesimals can lead to the enrichment of planets in heavier elements, i.e. to higher metallicity. The carbon enrichment in the solar giants has already been detected for Jupiter (three times the solar abundance), Uranus and Neptune (30 times that of solar abundance). Determining whether the majority of the EGPs are similarly enriched in heavy elements is one of the major goals of future observation and would greatly contribute to our understanding of planet formation and evolution.

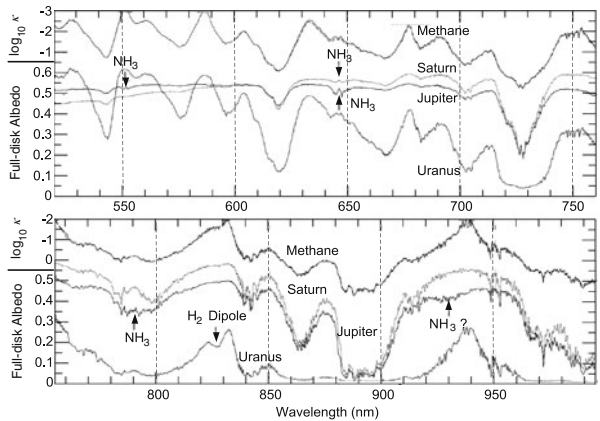

**Fig. 4** Full-disk albedo spectra of Jupiter, Saturn, and Uranus at 0.4 nm spectral resolution. For comparison of spectral features, the methane spectrum for the temperatures of the Jovian planets is shown on top with the *solid line* at 0.4 nm spectral resolution. The *dashed line* at 1 nm resolution is from Karkoschka [34]. Methane absorption is shown on a logarithmic scale. *Arrows mark* wavelengths of some ammonia bands and hydrogen dipole absorptions bands [34]

## 2.3 Targets

A spectropolarimetric instrument allows for the observation of inactive stars. Indeed, the light of active stars is usually sufficiently polarised to contaminate any usable signal from a planet. M stars being quite active, the targets must thus be limited to inactive F, G, and K stars. Moreover, the optical design of this proposal and its elliptical mirror would allow for the observation of targets up to 30 pc away while achieving reasonable integration times (see Figs. 8 and 9 for estimates of the integration times). These two properties of the target population (inactive and nearby stars) lead to a target sample which is very close to that of the radial velocity surveys (HARPS and Lick/Keck/AAT in the present, ESPRESSO and CODEX in the future).

Since an eight-octant phase mask (EOPM) is proposed as the coronagraph, the inner working angle of the instrument is equal to the angular resolution of the telescope. It translates into a linear relation between the minimal observable planetary semi-major axis $a_{min}$ around a target star at distance $D_*$: $a_{min} \propto D_*$. The spectropolarimetric technique yields an improvement in the contrast ratio between the planetary and stellar fluxes of five orders of magnitude; the EOPM improves the contrast by another ten orders of magnitudes. This means that planetary light with a flux ratio of up to $\phi_p/\phi_* > 10^{-11}$ can be distinguished. Using a very rough estimation of the stellar flux reflected by the disk of a planet of radius $R$, albedo 0.4 and semi-major axis $a$, seen at a phase angle of 90°, one can deduce that $\phi_p = 0.4 \times 0.5 \times (\pi R^2/4\pi a^2) \times \phi_*$. This gives a limiting relation between the observable planetary radii with respect to the semi-major axes:

$$\frac{a}{1\,\text{AU}} \lessapprox 3 \cdot \frac{R}{R_\oplus}. \tag{1}$$

This relation means that the achievable contrast using the combination of coronagraphic and spectropolarimetric techniques described in this paper allows, theoretically, for the observation of planets of 1 Earth radius up to 3 AU, and 10 Earth radii up to 30 AU. Unfortunately, much harsher restrictions follow from photon flux induced integration time limits (see Section 3.2.6) as well as the telescope's resolution capacities (see Section 3.2.2). Nevertheless, the higher the achievable contrast ratio, the better the conditions of delectability in terms of technical feasibility, e.g. CCD full-well capacity.

### 2.3.1 Statistics in the solar neighborhood

The Search telescope of this study would observe in a 45° cone always facing away from the sun because of the design of the telescope's sun shield. This constraint is also a part of the Darwin mission design (see [13]). As such, the Darwin star catalog can be used [31, 33], which provides a reasonable number of 280 target stars with spectral types suited to spectropolarimetry, as shown in Fig. 5. The current extrasolar planetary surveys have found planets around ≈ 6% of all observed stars. The growing number of ground and space based instruments as well as their improving sensitivities will increase this number.

Figure 6 (left) shows the planetary companions which have already been detected under 30 pc in a mass–semimajor axis diagram. The red rectangle

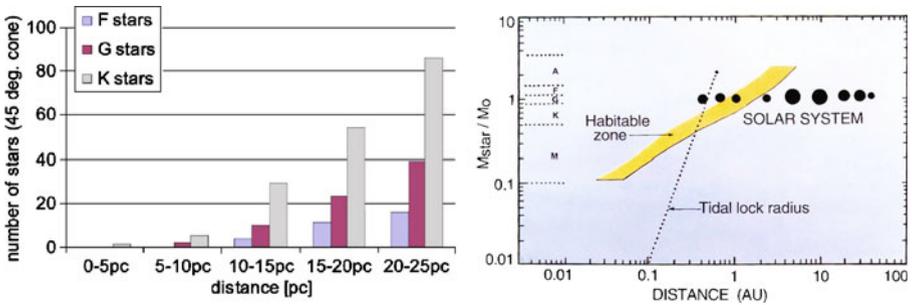

**Fig. 5** *Left* Number of F, G, K suitable target stars vs. distance in a 45° cone [31]. *Right* Habitable Zone as a function of stellar type and orbital distance [36]

shows the main interest target population, between 1 and 10 Earth masses and between 1 and 5 AU, while the black rectangle shows all the accessible targets. There are already a good number of accessible targets, and present and future radial velocity surveys are expected to provide a suitable number of planets in the primary interest range. Figure 6 (right) shows the same mass–semimajor axis diagram for all the already detected exoplanets without limitation on stellar distance, and shows that targets in the main interest domain are becoming accessible.

### 2.3.2 Example mission schedule

The proposed scientific schedule consists of the first third of the mission spent on an exploration of a target list as large as possible. This preliminary phase

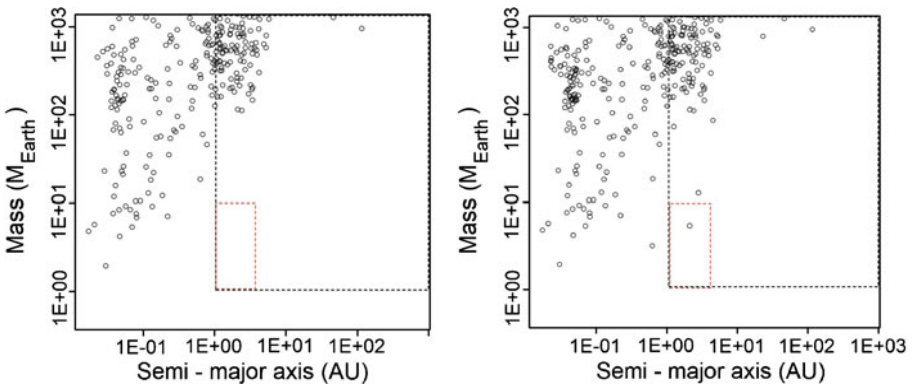

**Fig. 6** *Left* Mass vs semi major axis for all detected extrasolar planets around stars under 30 pc. *Right* Mass vs semi major axis for all detected extrasolar planets. Based on data from exoplanet.eu. The detections within the target box are achieved by gravitational microlensing

will be conducted using the low resolution mode observations, in order to obtain the broad features of the signal. Ideally, targets will be observed several times at different orbital and phase angles to derive an orbital inclination for them. After an analysis of the data gathered during this preliminary phase, the rest of the mission will be dedicated to in-depth investigation of the most interesting targets, using the high resolution mode of the spectropolarimeter. Amongst other properties this may allow the detection of the presence or absence of potential biomarkers as pointed out in Table 1. In the final phases of the mission, the instrument could be opened up to a more general scientific community. It could for instance be put to use in the study of protoplanetary disks, multiple stars, highly polarised objects such as AGNs, etc.

## 3 Payload design

### 3.1 Polarimetry concept

As outlined in the science case spectropolarimetry has significant potential in the characterisation and indeed the detection of exoplanets. The power of polarisation for characterising planets has been demonstrated numerous times in past studies of Solar System planets, such as Venus, Mars, Jupiter and Earth, see Hansen and Hovenier [25], Braak et al. [11] and Ebisawa and Dollfus [17].

The technique involves measuring the Stokes $Q$ and $U$ parameters, which are essentially determined as the difference between the intensity of orthogonally polarised components of the electric field. The $Q$ parameter is found when one axis lies in the plane of polarisation and $U$ when the system is rotated through 45°. The exact system configuration is outlined in the following sections.

### 3.2 Optical system

#### 3.2.1 Main optical system

To meet the measurement criteria stated above an optical telescope is proposed in Cassegrain layout with an f/1 elliptical mirror with a size of $9 \times 3.7$ m. The secondary mirror is placed on-axis and has roughly 1/10 the size of the primary. Including the obstruction of the secondary the aperture size is $\pi \cdot 9 \cdot 3.7 - \pi \cdot 0.9 \cdot 1.85 \approx 25.5$ m$^2$. The mirror size is a tradeoff between angular resolution, photon flux and the present launch capabilities of an Ariane 5 ECB launcher. To fit the telescope into the launcher the main mirror will be made of two segments. This will call for high accuracy positioning in the alignment of the two segments. This setup is again a tradeoff between the optical quality and the complexity of the deployment mechanisms needed for the mirror and sun shield.

*3.2.2 Angular resolution*

To resolve a planet 1 AU distant to its host star up to a distance of 30 pc an angular resolution of 1/30 arcsec = 0.033 arcsec is required. The angular resolution of a telescope is determined by applying the Rayleigh-Criterion

$$\beta_{\min} = 1.22\frac{\lambda}{D} \tag{2}$$

where $\lambda$ is the wavelength and $D$ is the diameter of telescope aperture.

Thus a mirror with a diameter of 9 m has an angular resolution of $\beta_{\min} = 1.22\frac{900 \cdot 10^{-9}}{9} = 0.025$ arcsec at 900 nm. This would be enough to resolve the planet at the longest wavelength considered for this proposal's observations. But the resolving power of the whole optical system depends on several factors. Since the contrast ratio between the star and the planet is too large to directly record the signal the use of a coronagraph is inevitable. As is explained in Section 3.3 these instruments have a certain inner working angle (IWA). The IWA of the coronagraph used in this mission will be between 1.0 and 2.0 $\lambda/D$. As such, in the worst case scenario the proposed 9-m telescope would only be capable of separating a true Earth-like planet (1 AU from the host star) in a distance of 30 pc at a wavelength of approximately 600 nm. This would restrict the mission at the outer edge of the 30-pc observation radius. But as explained in Section 2.2 this would only result in a minor drawback to the scientific goals of the mission. The goal is not only to characterise true Earth-like planets but also to shed light on the diversity of extrasolar planets in the low mass range.

*3.2.3 Quality of the main optical system*

Aside from the requirement for high angular resolution, the high contrast ratio between star and planet proves the main optical challenge. In theory the optical system images all the collected light from a point source in an Airy disc which is the Fourier transform of the aperture geometry. In a real optical system the imperfect optical surfaces always reflect light on the area outside the Airy disc. In this area the light interferes and creates a characteristic speckle pattern (see Section 3.2.4). A main criterion for the quality of an optical system is the Strehl-ratio $S$. This is the ratio between the intensity collected within the real Airy disc and the theoretical intensity collected in an ideal Airy disc. For highly corrected optics, the Strehl-ratio of a mirror can directly be calculated from the standard deviation of the mirror surface:

$$S = e^{-\sigma^2} \approx 1 - \sigma^2 \tag{3}$$

Here $\sigma$ is the standard deviation of the mirror surface normalized on the observation wavelength. The approximation is valid for small standard deviations. To make the Strehl-ratio as high as possible the mirror surface has to be very smooth. Considering the large mirror of the SEARCH design, one cannot expect a low standard deviation over the whole mirror surface. For this reason active optics (AO) actuators are introduced, which can correct the surface variations of the mirror in place. ("Active optics" may be distinguished from the "adaptive optics" systems used to correct atmospheric distortions). That means the mirror has only to be manufactured with high precision over patch sizes corresponding to the displacement between the actuators.

### 3.2.4 Reduction of speckle noise

Given a telescope with a certain Strehl ratio $S$ a fraction $(1 - S)$ of the light entering the system is not focused into the Airy function, but spread over the image plane in a characteristic speckle pattern. A speckle is a patch of light with the size of an Airy function which—in this case—becomes especially problematic, because it resembles the image of a planet. According to Bloemhof [6] and Roddier [63] two types of speckle patterns are created: an antisymmetric and a symmetric pattern. The antisymmetric speckles are located on the Airy disc and are effectively removed by the coronagraph, the symmetric pattern however is not affected by the coronagraph. The relative intensity of a quadratic speckle behind the coronagraph and the Lyot stop is approximately given by

$$I_{\text{quadratic}} \approx (1 - S) \frac{1}{0.352} \left( \frac{a}{D} \right)^2 \left( \frac{D_{\text{Lyot}}}{D} \right)^2 . \tag{4}$$

where $S$ is the Strehl-ratio, $a$ is a characteristic coefficient, $D$ is the aperture diameter and $D_{\text{Lyot}}$ is the diameter of the Lyot stop projected into the aperture.

The characteristic coefficient is approximately given by the displacement of the Active Optics (AO) actuators. The intensity is normalised on the Strehl ratio $S$ and therefore gives directly the mean contrast between the star and a speckle. In Fig. 7 (right) the relative speckle intensity is calculated using (3) and (4), and plotted for different surface standard deviations (Strehl ratios) and Lyot stops. It is obvious that neither a smaller surface standard deviation nor a larger Lyot stop will decrease the speckle intensity significantly. According to Fig. 7 (right) the speckle intensity can only be significantly lowered with AO actuator displacements of less than 5 cm on the primary mirror or less than 5 mm on the secondary. A high AO actuator density does represent a technical challenge, nonetheless, it would prove more practical to place high-density AO on the secondary mirror due to size and weight considerations. A surface standard deviation of $\sigma = 1/150$ yielding a Strehl ratio of $S = 0.993$, a medium distance of $a = 0.01$ m between the AO actuators, a primary mirror diameter of $D = 9.5$ m and a moderate Lyot stop of 8 m would lead to a speckle intensity of about $10^{-8}$. This is one to three orders of magnitude too low to directly image an exoplanet. Therefore, additional measures for speckle reduction are inevitable. One possibility would be to close the Lyot stop even more, but this would directly affect integration times. If the dynamic range of the CCD-chip is large enough, the symmetry of the speckle pattern could be used to effectively suppress it after the recording. In such a case, the image would be copied, flipped by 180° and subtracted from the original. With the speckle pattern being symmetrical the only asymmetric part in the image would be the planet. This method can also be applied to spectroscopy if an integral field spectrograph is used.

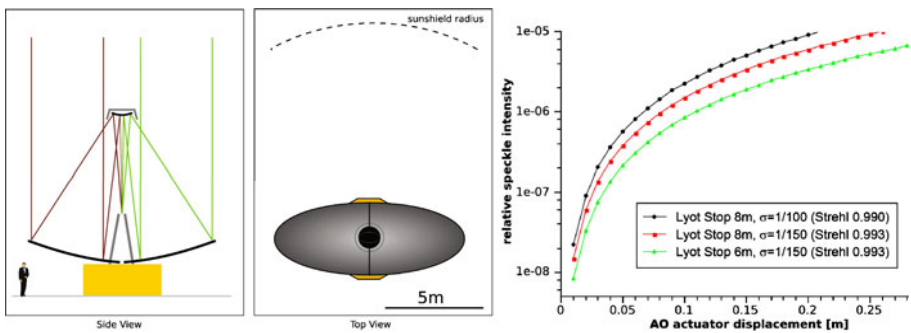

**Fig. 7** *Left* Telescopic system. *Right* Calculation of relative speckle intensity for different systems

A very elegant way to get rid of speckles is interferometric speckle nulling. This technique is discussed in Bordé and Traub [9, 10] and Guyon [22]. Speckle nulling has to be considered carefully. It is a very powerful, which could prove necessary to directly image faint exoplanets. Of course this presents several technical issues to be overcome, including a certain loss of light, which is critical when working with a very limited aperture size. The speckle noise level also leads to very strict requirements to the accuracy of the polarisation measurements. To measure a polarisation signal to an accuracy of 0.5% with speckle noise being three orders of magnitude higher than the signal itself, an on-axis telescope with very small cross-polarisation is required.

### 3.2.5 Straylight considerations

There are different sources of straylight to be taken into account. Since the intention is to detect the very faint signal of the planet very close to its host star special measures are required to reduce the stray light. Primary sources of straylight include:

1. light emitted from the sun and the Earth–Moon system
2. light scattered from the spacecraft
3. local zodiacal light

Overcoming the **first** point is accomplished using a deployable sunshield and always pointing the spacecraft away from the sun. This allows an observation angle of ±45° above and beneath the ecliptic.

The amount of stray light generated by obstructions in the optical path (**second** point) depends on the spacecraft design, which is discussed later in Section 4. The boom holding the secondary mirror is placed in the middle of the primary mirror as shown in Fig. 13. As such, the struts holding the secondary mirror are obstructing the aperture. In this situation the image of the star would have additional spikes extending outwards from its center. These spikes point $\pm 45°$ away from the major axis of the primary mirror. Thus, the line of highest resolution parallel to the major axis will be free of this straylight. The **second** point is taken care of using a tapered baffle and a pinhole near to the first focal plane of primary and secondary mirror. This system can reduce the stray light coming from other light sources outside the aperture. For a very powerful baffle design see Plesseria et al. [60].

The **third** point is not a major problem in this design because it only delivers a constant intensity background to the observations which can be subtracted during the image post processing as shown in Section 3.2.4.

### 3.2.6 Photon flux

The expected count rate (*cnt* [photons/s]) of the observed target has been calculated as follows:

$$cnt = \frac{\pi}{2} \cdot \left( \frac{r_\star r_p}{d_\star d_p} \right)^2 \cdot \int_\lambda^{\lambda + \Delta\lambda} \frac{I(T, \lambda)}{E_{\text{photon}}(\lambda)} d\lambda \cdot \phi \cdot a_{\text{geo}} \cdot A \cdot \Theta \cdot Q \qquad (5)$$

Here $r_\star$, $d_\star$ denote the radius and the distance of the host-star respectively, $r_p$, $d_p$ the target planet's radius, and the distance to its host star. The black body irradiation intensity function per photon energy $\frac{I(T,\lambda)}{E_{\text{photon}}(\lambda)}$ is integrated over spectral bands,[1] corresponding to the spectrograph's resolution (e.g. $R = 70 \rightarrow \Delta\lambda \simeq 4.3$ nm at $\lambda = 300$ nm), neglecting stellar and planetary absorption losses. $\phi$ represents the orbital phase factor, ranging from 1 at full phase (phase angle $\alpha = \pi$) to 0 at $\alpha = 0$. The planet's geometric albedo $a_{\text{geo}}$, the primary mirror's surface $A$, the total instrumental throughput $\Theta$, as well as the detector's quantum efficiency $Q$ range from 0 to 1. The count rate is linked to the required signal to noise ratio ($SNR$) via

$$cnt = \frac{SNR}{\tau} \left( SNR + \sqrt{4N_{\text{pix}} \cdot (n_{\text{readout}}^2 + n_{\text{straylight}} + n_{\text{thermal}}) + SNR^2} \right) \quad (6)$$

Where $\tau$ denotes total exposure time, $N_{\text{pix}}$ the number of CCD-pixels, $n_{\text{readout}}$ the number of photons due to read-out, $n_{\text{straylight}}$ due to stray-light and $n_{\text{thermal}}$ due to thermal noise. Combining (5) and (6), one can derive an expression for the systems range in terms of planetary radii.

$$r_p = \frac{d_p d_\star}{r_\star} \sqrt{\frac{2}{\pi} \frac{SNR^2 + SNR\sqrt{4N_{\text{pix}} \cdot (n_{\text{readout}}^2 + n_{\text{straylight}} + n_{\text{thermal}}) + SNR^2}}{\tau \cdot \phi \cdot a_{\text{geo}} \cdot A \cdot \theta \cdot Q \cdot \int_\lambda^{\lambda + \Delta\lambda} \frac{I(T,\lambda)}{E_{\text{photon}}(\lambda)} d\lambda}}$$

$$(7)$$

---

[1]The numerical integration of Planck's function has been performed following Widger and Woodall [81].

Present assumptions on the setup proposed are $\Theta = 0.3$,[2] $a_{geo} = a_{Earth} = 0.36$, and the measurements are considered at $\phi = 0.5$. The two segments of the primary mirror have a total collecting area of $A = 25$ m$^2$. Since no specific assumptions on the detection device have been made, and no reliable stray-light calculations could be performed until further investigations into the main optical design can take place, $N_{pix}$ has been set to zero, driving SNR calculation towards pure photon noise. Nevertheless, a reasonable quantum-efficiency of $Q = 0.9$ over the whole spectral band has been assumed. Stellar radii have been interpolated from Habets and Heinze [23] corresponding to F-K type surface temperatures.

The resulting estimates on the instrumental range can be regarded in Fig. 8. Assuming a total integration time of one day per spectrum, a characterisation of Earth like exoplanets around G-stars and Super Earths around fainter stars is possible providing a spectral resolution of 70. Compared to the restrictions on continuous observation time due to the spacecraft's solar panel configuration—currently approximately 90 days—polarimetry can also be performed. The maximum exposure plots (45 days, Fig. 9) were generated having a very generous amount of time reserved for down-link-, re-calibration and reorientation phases of the spacecraft. It is plainly visible, that high resolution spectra can be achieved for planets within the habitable zone around G-type stars for a slightly narrowed spectral range at a distance of 10 pc. At 30-pc high resolution spectrometry can still be performed for sub-Jovian planets.

## 3.3 Instrument

The optical layout of the instrument is outlined in Fig. 11. A field stop is positioned at the focal point of the telescope, through which the light enters the instrument. A lens then refocuses the light onto a coronagraph. This serves the purpose of blocking out the host star light of the planet under investigation. The subsequent optics includes a collimator, Lyot stop and the fibre bundle placed at the imaging plane.

## 3.4 Coronagraph

The specific type of coronagraph selected for this project is the Eight Octant Phase Mask or EOPM [56]. The mask is composed of eight sections, each of which is phase-shifted by $\pi$ compared to its neighbouring octant (Fig. 10). Thus, any symmetrical image centered on the mask undergoes destructive interference and is cancelled out. The light focused by the first lens of the system produces an Airy disk centered exactly on the crosshair of the phase mask. Due to the design of the mask, destructive interference occurs between

---

[2]Regarding this estimate, the phase mask coronagraph will reduce the planets light close to the optical axis by approximately 50%, as can be seen in Murakami et al. [55]. The photonic spectrograph has been considered as having a throughput of 60%. All other components are considered as providing lossless performance, except for the CCD's quantum efficiency.

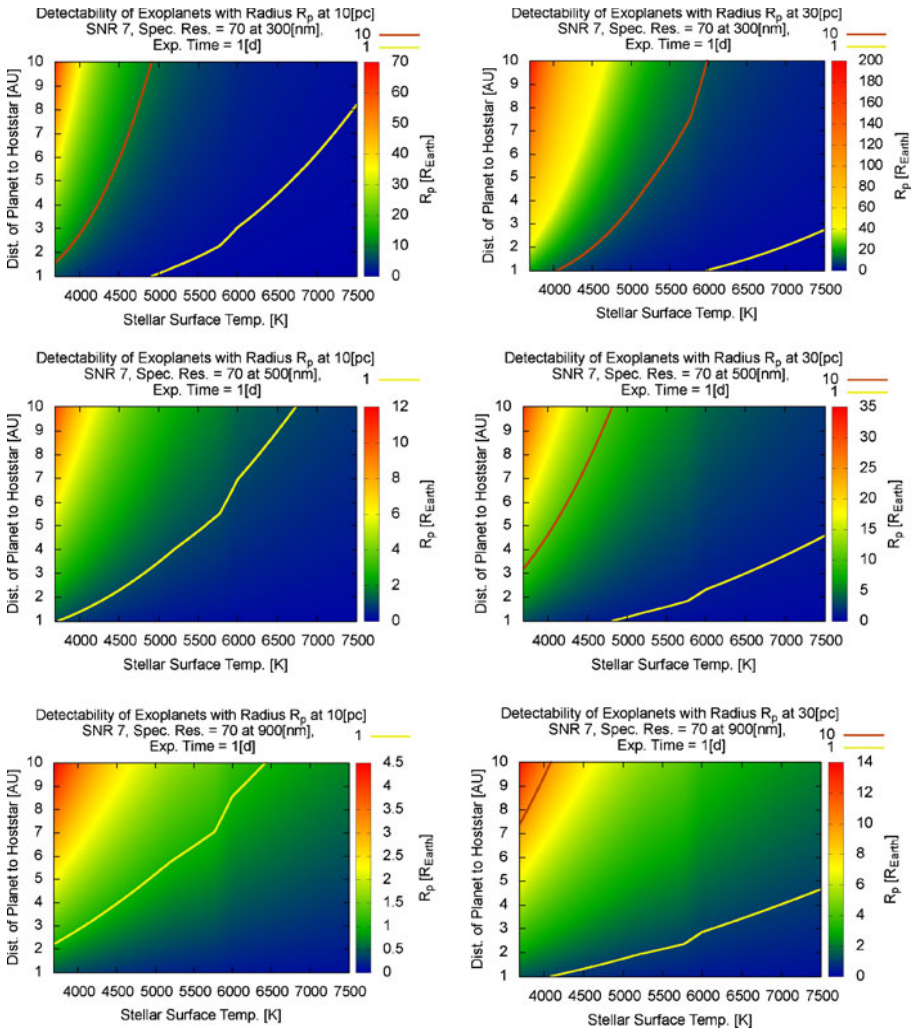

**Fig. 8** Instrumental range in low resolution mode. *Left* Spectrometric capacities at a distance of 10 pc. *Right* At a distance of 30 pc. Terrestrial planets around G-stars can be fully characterised almost up to 30 pc

the phase shifted components of the electric field. As such, at the next pupil plane most of the light is in the area surrounding the pupil. A Lyot stop in this pupil plane blocks out the light surrounding the pupil. Thus, when focusing the electric field after the Lyot stop the flux in the center is attenuated significantly. Optimum attenuation occurs only for an object perfectly centered on the mask's crosshair, any off-axis object or structure in the nearby region of a star (such as an orbiting planet) will not suffer this effect and remain visible. A more detailed explanation of the function of phase mask technology in general can be found in Murakami et al. [56]. One clear challenge in the application

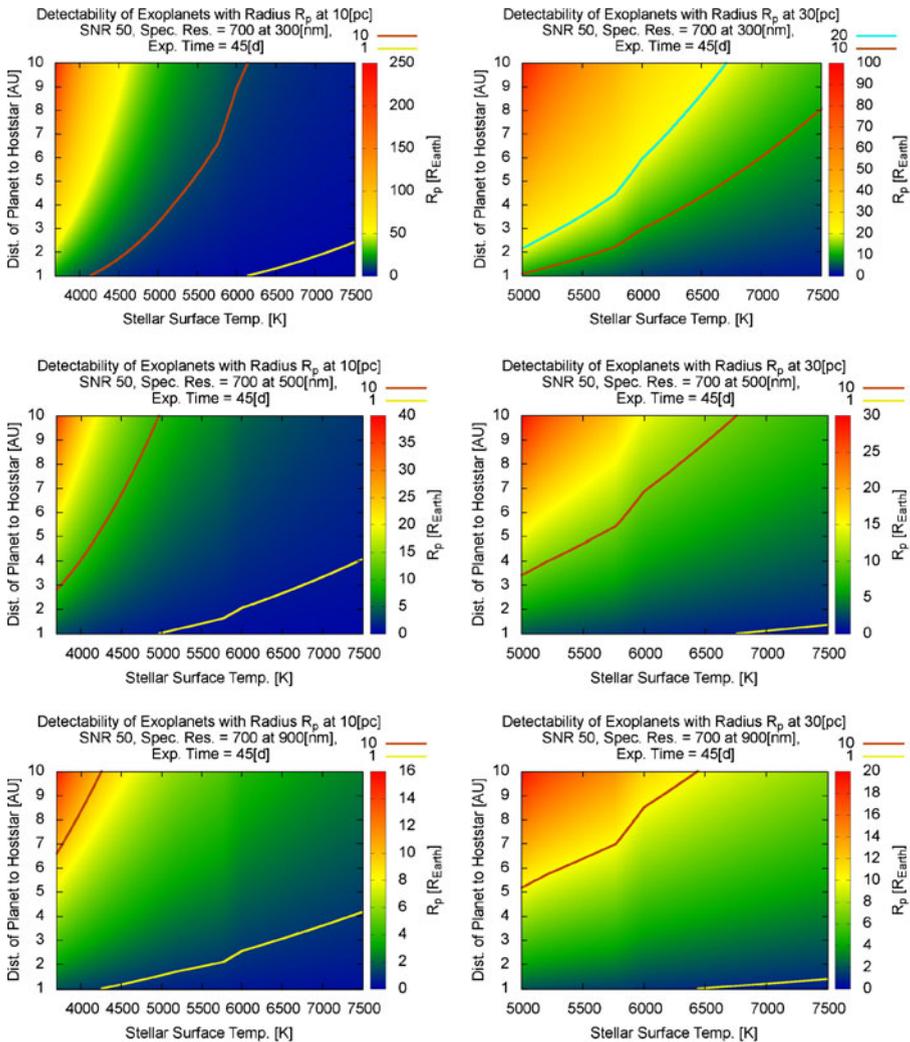

**Fig. 9** Instrumental range at maximum exposure time, high resolution mode. *Left* Spectrometric capacities at a distance of 10 pc. *Right* At a distance of 30 pc. Super Earths around G-stars may be characterised with a spectral resolution of 700 up to 30 pc

of the phase mask technology is the large spectral bandwidth of 300–900 nm considered by the mission. Achromatic development in this technology should prove capable of meeting these requirements. Approaches for achieving achromaticity include multi-layer thin film design or an emulation of dielectric phase plates used as achromatic phase shifters in nulling interferometers [7]. More details on the development of achromatic phase masks can be found in Boccaletti et al. [8], which outlines a prototype covering a spectral range of 950–1,800 nm.

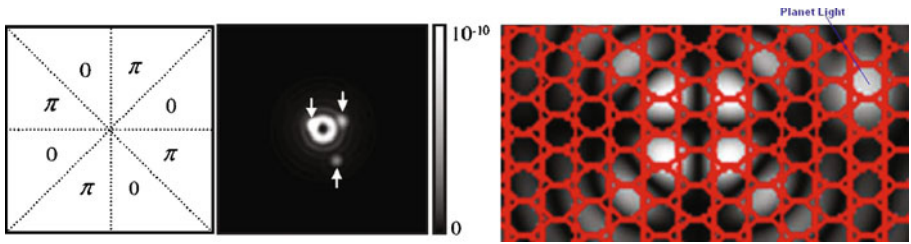

**Fig. 10** *Left* Schematic of the EOPM mask [56]. *Right* Graphic representation of a nulled stellar image resulting from a phase mask with an overlay of the fibre arrangement, each one of which feeds back to a single spectrograph

Although the chromaticity of the coronagraph is not expected to pose a significant problem, the possibility exists of mounting two EOPMs with differing central frequencies onto a rotator wheel and shifting between them. Further restrictions in the spectral range may result from the other optical components, such as for example the lens materials. Although this may prevent the lower limit of 300 nm being reached, a lower limit of 400 nm would not prove problematic. The EOPM technology has to date been prototyped using liquid crystals. Should the technology not have advanced beyond the requirement for liquid crystals, which are not yet space proofed, then the Four Quadrant Phase Mask could be employed in its place [56].

## 3.5 Integral field spectrograph

### 3.5.1 Imaging outline

The coronagraphic mask reduces the stellar light by a factor of up to $10^{-11}$. Despite the significant reduction in star light the planet light will still be much fainter than that of the star. In order to extract the spectrum of the planet, the point spread function (PSF) is sampled over 800 fibres, each of which feeds back to an individual spectrometer. As such, the star light is distributed over the entire array, whereas the planet light is contained across a small bundle of fibres.

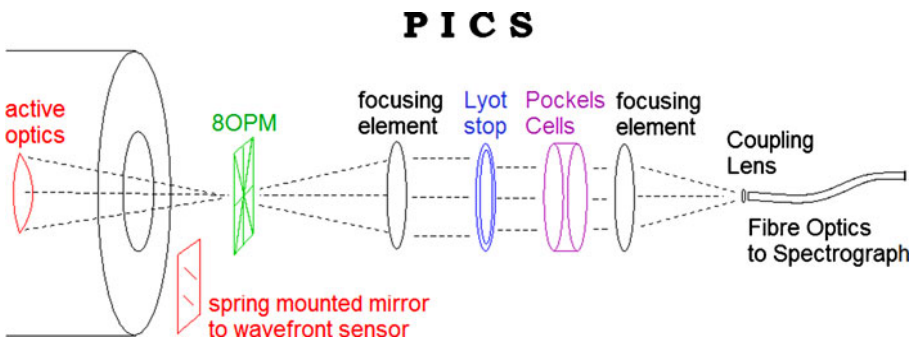

**Fig. 11** Outline of the optical path of the instrument

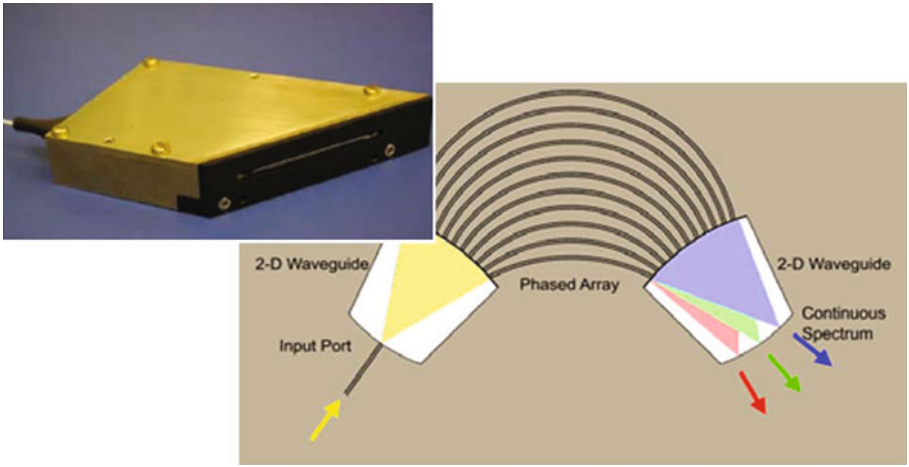

**Fig. 12** The Integrated Photonics Spectrograph

### 3.5.2 Spectrographs

Considering that each of the 800 fibres feeds back to a separate spectrometer, a particularly light weight and affordable unit is required. Meeting both of these requirements is the Integrated Photonics Spectrograph from the Anglo-Australian Observatory. Each of these fibre spectroscopes have dimensions of 100 mm × 75 mm × 2 mm, a mass of 0.8 kg and a transmission profile of 60–65%, see Fig. 12. This results in a total array dimension of 1.2 m × 1.5 m × 0.015 m and mass of 630 kg.

### 3.6 Polarimetry

As previously stated spectropolarimetry is the technique chosen to accomplish the science goals outlined. Apart from yielding polarimetric data, which is of itself scientifically important, polarimetry also improves the contrast ratio between the star and the planet. The technique involves examining the difference between P and S polarised light to determine the Stokes Q and U vectors. This can be done either temporally or spatially. The spatial approach employs a polarising beam splitter and reads in both the P and S polarised beams simultaneously on two different CCDs. A disadvantage of this approach is the calibration requirement of the CCDs, which must be as identical as possible to ensure that only the difference in radiation intensity is measured. As such, the temporal option is favoured in this study (this approach is adopted in the ZIMPOL device [66]).

The same CCD design as implemented in ZIMPOL is also proposed for use here. This requires that every second row of the CCD be masked so that charge packages created in the unmasked row during one half of the modulation cycle are shifted for the second half of the cycle to the next masked row, which is used as temporary buffer storage. After several thousand modulation periods the CCD is read out in less than 1 s. The sum of the two images is proportional to the intensity, while the normalised difference is the polarisation degree of one Stokes component [66]. Pockels cells are suggested as polarisers, these would have high transmission of over 98% and avoid the necessity of rotating parts. Two Pockels cells could be placed in the collimated beam after the phase mask. These cells would be oriented orthogonal to one another. As such, when a voltage was applied to one it would polarise at 90° whereas the second cell would polarise at 0° when subjected to the same voltage. The Pockels cells could then be activated and deactivated as required to polarise the signal. Should the use of Pockels cells prove in any way problematic the option of using standard polarisers on a rotator wheel could also be used.

### 3.7 Measurement procedure

Before the actual image acquisition the optical system is calibrated to ensure maximum suppression of host starlight and speckles. First of all the coronagraph is aligned with the help of a wavefront sensor which monitors the light from the host star. The actuators on the telescope mirrors are iteratively adjusted in such a way that the light from the host star is minimised behind the coronagraph. The effect of the coronagraph can be measured by a CCD-camera or a photometer. Because the host star is usually very bright, only a very small fraction of the incoming light is needed for both wavefront sensing and photometric measurements. That means alignment of the coronagraph can be measured and maintained during image acquisition.

If the telescope moves thermally more than about one quarter of a wavelength the speckle pattern changes. Therefore, either the telescope has to be stable enough to ensure speckle nulling over the time of image acquisition or speckle nulling has to be done in real time during the measurement. The latter would of course result in greater loss of light. In this way a stack of images is recorded. The exposure time for each individual image has to be small enough that the remaining light from the host star does not overflow the CCD-detector. This procedure is valid for both direct imaging and integral-field-spectroscopy. In both cases a symmetric speckle nulling can be applied in post-processing. For polarimetry, different images are recorded with orthogonal states of the polariser.

## 4 Spacecraft

### 4.1 Spacecraft general design

The overall spacecraft design is driven by the telescope size and accommodation and the need for a very stable environment to achieve the scientific goals.

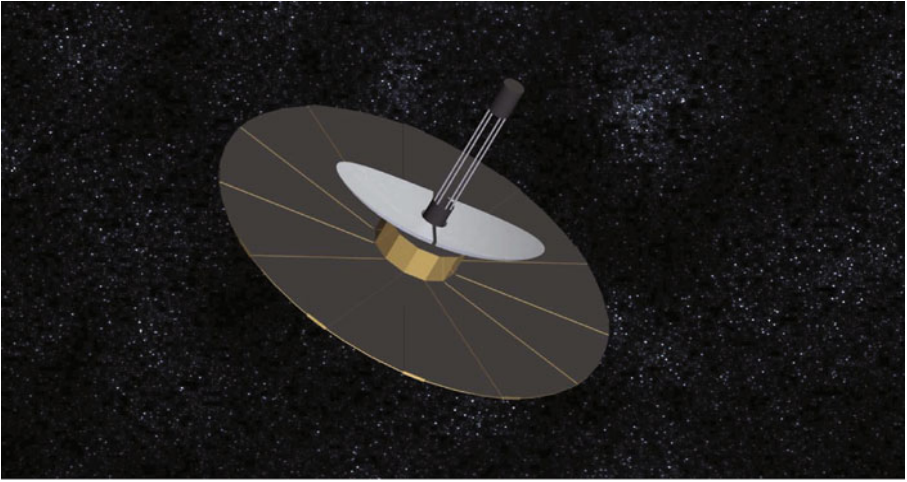

**Fig. 13** Search deployed configuration

The proposed spacecraft in its deployed configuration will be 9.5 m in height and 22 m in diameter for a total mass of 5.5 t. The primary mirror form an elliptical shape and surrounds the central boom, this is complemented by a 80-cm secondary mirror (also split in the center), mounted with active optics and located on top of a central boom of 8 m length from the primary mirrors position. They will be protected by a corona sunshield with a corona radius of 9 m, with a hexagonal shape, using the same technology as the GAIA project [21]. The deployed configuration can be seen on Fig. 13.

The critical parts of the spacecraft design include the general configuration, with particular emphasis placed on the sunshield to reduce stray light, and the deployment mechanisms. The sunshield is designed to protect the secondary mirror from direct sun light. As the spacecraft will point to different targets, it must be shielded within a pointing range of [−45°, 45°] making 14% of the sky available for observation at a given time. These values assume the sunshield is fixed at a 9-m radius and with a hexagonal configuration. During the launch phase, the mirrors and the sunshield are in a stowed configuration, as seen in Fig. 14 (right).

In space, the circular sunshield is deployed to its full extent in order to shield the two off-axis mirrors. Moreover, due to the vibrations expected during launch, the secondary mirror boom needs to be in a stowed configuration of 7.5 m length. The boom is then extended in space with a telescopic system to the required length of 9.5 m. The mission could be launched by an Ariane 5 ECB from the Guiana Space Center at Kourou. It would then be transferred to an orbit around the Lagrange L2 point of the Sun–Earth system with a cruise duration of 3 months. The mission baseline would be 5 years in orbit once it arrived and had successfully completed its commissioning phase.

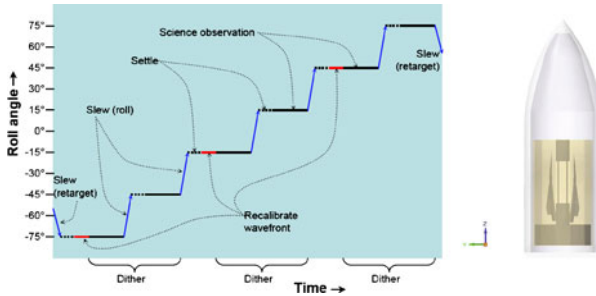

**Fig. 14** *Left* SEARCH possible mission operation scenario for the first year: timeline for a planet observation around one star. The observation includes six science integration periods separated by five roll slews. Each slew is followed by settling time. Each pair of science integrations constitutes a dither, and each dither is preceded by recalibration of the wavefront [47]. *Right* SEARCH stowed configuration

The mission observation scenario would be as follows: The maximum time of exposure for the selected targets is approximately 1 day for low-resolution scans. Since the orbital inclination of the target will not generally be available, a period of quick observations of the target at different spacecraft roll angles would first be carried out. These would follow the TPF-C observation strategy at six roll angles, as illustrated in Fig. 14. Assuming a day for manoeuvres and data transmission each month, it would be possible to achieve this general sweep of the selected targets during the first year of the mission, however very little time would remain for other observations. The second phase of the mission would focus on more precise observations of interesting targets. Assuming an exposure time of 45 days for high-resolution spectroscopy (including manoeuvres and data transmission), it would be possible to have a precise observation for two out of five targets, and still have 50% of the observation time available for general astronomical observations. Alternatively, a dedicated mission could achieve detailed observations of up to four out of five targets.

## 4.2 Spacecraft subsystems

The Attitude and Orbit Control System (AOCS) is one of the most critical parts of the spacecraft's subsystems, because of high pointing and stability requirements of the optics. The main requirements are 0.01 arcsec of pointing accuracy and a maximum of 3 ma of deviation during the typical integration time of 1 day for low-resolution images, and more for high-resolution scans, with a total integration time of up to 45 days. To achieve this, both star trackers and gyrometers will be used to make a first measure of position. CCD data from the telescope itself will then be made directly available to the AOCS loop, and 16 FEEP thrusters in order to counter the very small perturbations of the L2 environment. 12 classical thrusters each of 10 N will be used for orbit correction manoeuvres.

The spacecraft is protected from solar irradiation by the deployable sun-shield covered with multi layer insulation sheets. Where required, heaters and multi layer insulation are used at other parts of the spacecraft, resulting in a mostly passive thermal control. In order to guarantee thermal stability at the sensors, the CCD will be passively cooled by means of a heat pipe and a radiator.

The scientific data will be transmitted in X-band through a set of phased array antennae located at the bottom of the spacecraft. This high gain antenna concept is based on the GAIA communication architecture. Two low-gain omni-directional antennas are used for telemetry and telecommand, available also during eclipses or other survival situations. In order to meet the power requirements the spacecraft will have 13 m$^2$ of solar panels, with a Li-ion battery system providing power during critical phases. The solar arrays geometry is similar to the GAIA configuration, with six separate solar panels in a hexagonal arrangement.

### 4.3 Possible mission schedule

Figure 15 depicts a possible schedule of the mission. Accounting for necessary technology development phases A and B have been extended. Launch would then be foreseen by 2027. The scientific duration of the mission would be 5 years; extendable by 2 years depending on the fuel reserves.

### 4.4 TRL and risk analysis

The Technology Readiness Level (TRL) is a measure used by space agencies to assess the feasibility of a given technology. It has a scale ranging from 1 to 9, 1 being "Basic principles observed and reported" and 9 "Actual system Flight proven through successful mission operations" according to ESA's definition of TRL [18].

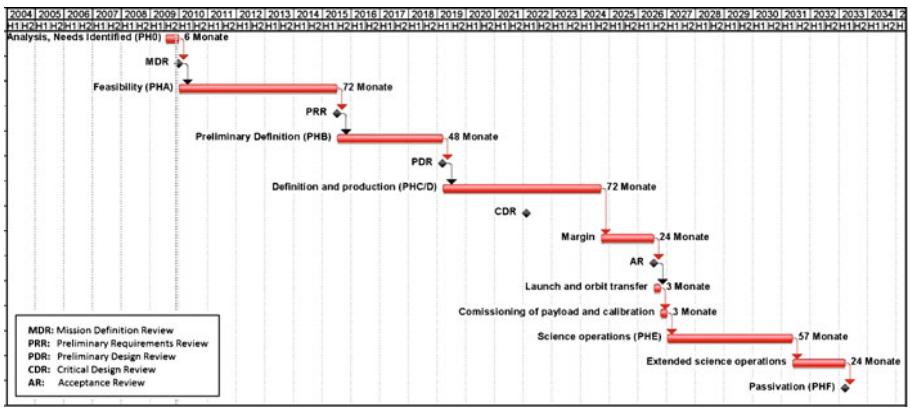

**Fig. 15** Search—Possible mission schedule

No detailed analysis of the readiness of the various subsystems has yet been carried out. Further work is needed to define a clear roadmap of technology development and a risk assessment. However, some critical technologies required in order to satisfy the scientific requirements can be highlighted. The grinding of the mirror will prove a major challenge, as well as the development of a set of new deployment mechanisms for the main mirror and deployable boom. Other subsystems, such as the sunshield and AOCS thrusters, have already been developed for missions such as LISA Pathfinder and GAIA but will require some modifications. The main driver, as is often the case in the design of a satellite system, will be the mass of the spacecraft, which needs to satisfy the capability of a launch on Ariane 5 ME with an injection into an orbit around the Sun–Earth Langrange point L2.

## 4.5 Descoping and upscoping options

The presented configuration may be considered the baseline design. The mirror size is one of the main issues: the Ariane fairing size is not a limitation, so it is possible to increase the mirror size if more ambitious scientific goals are considered, or decrease it if the mission proves itself too expensive. It is also possible to descope the mission to only one mirror and to abandon the deployment challenges, however this would greatly decrease the mission performance.

### 4.5.1 Scientific consequences

As discussed above, the primary mirror of the proposed system is the main driver in the cost of the mission. On the other hand, the ambitious scientific goals of characterising Earth-like planets require such a large effective diameter—as described in Section 3.2. The proposed system with a mirror divided in to two 4-m segments is considered a reasonable trade-off between meeting the scientific goals and the expense and technology development. Descoping the total mirror size from 9 to 4 m (Fig. 16) would significantly limit the possible targets. The limitation is a result of the decreased resolution, and would mean that SEARCH could only observe planets within a minimum distance of 3 AU to the host star at 30 pc.

Consequently Earth size planets could not remain in the target list, and SEARCH would have to focus on characterising the diversities of gaseous planets (Section 2.2.2). Receiving this kind of information would still be of great interest in understanding the atmospheric composition and evolution of the outer regions of planetary systems. However, the even more interesting inner regions would not be accessible. It is also possible to lower the cost of the mission by simply using one half of the proposed system (only one primary mirror) as an off axes telescope as proposed for the Japanese Terrestrial Planet Finder(JTPF), with the same targets limitations.

**Fig. 16** SEARCH upscoping and descoping options. The upscoping option has a 20-m mirror size and the descoping a 4-m size

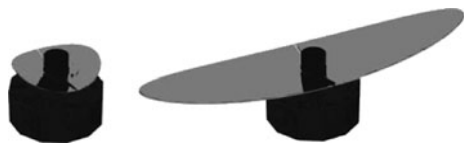

The most significant technical challenge is seen to be in the production of such large mirrors. As shown, the major axes of the mirror could be decreased in size or increased up to 20 m, whereas the minor axes will be the same for all designs, so the manufacturing of such a mirror will prove a major issue. It is also possible to envisage segmenting either of the two 10-m mirrors. The proposed launch and deployment strategies should be possible with an Ariane 5. Due to the symmetric deployment, problems with the center of gravity being off axes in the launch configuration can be overcome (see Fig. 16).

## 5 Summary and discussion

The dedicated space mission SEARCH (**S**pectropolarimetric **E**xtrasolar **A**tmosphe**R**e **CH**aracterisation), proposed in this paper, will study the diversity of terrestrial extra solar planets using spectropolarimetry. Since the host star's light scattered from an exoplanet will be polarised to a certain degree, depending on surface and atmospheric features, polarimetric measurements contain valuable information about these characteristics not accessible by other methods. Examples are cloud coverage, cloud particle size and shape, ocean coverage, atmospheric pressure among others. Also the change in the degree of polarisation over one orbital period contains valuable information on the orbital inclination, which is not accessible via radial velocity measurements. By combining this method with a spectral measurement, it is possible to information about the atmospheric composition and to detect molecules such as $H_2O$, $O_2$, $O_3$, $NH_3$ and $CH_4$. In order to achieve these ambitious goals two main obstacles have to be overcome: these are high contrast ratio between the planet and its host star and the need to spatially resolve those two with the spacecraft's telescope. To meet these two challenges, new design concepts have been developed and presented in this paper.

The proposed mirror is an f/1 parabolic mirror with elliptical rim of size of 9 m × 3.7 m in a Cassegrain layout. It is cut in half and folded towards the secondary mirror to fit into the Ariane 5 fairing. Thus, problems regarding the symmetric distribution of weight at launch can be overcome. Also an on axis-configuration significantly decreases the polarisation losses introduced by the system. This design allows for an angular resolution high enough to resolve planets as close to their host stars as 0.5 AU in a distance of 10 pc and as close as 1.4 AU in a distance of 30 pc, for the first time being able to characterise real Earth type planets in a significant target sample. It is also important to mention, that the proposed mirror design is a way to launch a mirror into space with Ariane 5 more or less regardless of its size in one dimension. If the technical problems in the manufacturing of such large mirrors can be overcome, launching a 20-m × 3.7-m mirror within Ariane 5 can be envisaged. The mirror's surface smoothness is critical to the quality of the optical system.

Since it is not possible to meet the required smoothness over the whole area of the mirrors, the concept of active optics has been proposed to correct the wavefront from deviations on the mirror surface.

Regarding contrast ratio, it has been shown that polarimetry itself can reduce the contrast ratio between host star and planet by five orders of magnitude during the post observational reduction phase, because the host star's light is on average unpolarised. For the proposed mission, a Half-Wave-Achromatic Eight Octant Phase Mask has been chosen as the coronograph and Pockel Cells suggested as polarisors. In addition, the use of an integral field spectrograph is proposed for separation of light from the star and the planet as well as the reduction of speckle noise.

Of course there are technological developments to be undertaken before a concept like the proposed SEARCH mission can be realised. Nevertheless, since nearly all of those are key developments for other planned or already scheduled space missions, future projects like SEARCH will be able to use these developments to achieve the ambitious goal of characterising Earth-like planets.

**Acknowledgements** The authors want to thank Andre Balogh, Antonio Castro, Malcolm Fridlund, Eike Günther, Günter Kargl, Helmut Lammer, and Jörg Weingrill for their valuable comments and useful discussions and Michaela Gitsch (FFG) for her organisatorial skills. This work originated at the Alpbach Summer School 2009; we thank the lecturers and tutors, as well as our fellow students for the unique learning experience there. The mission proposal presented in this paper was continued during the Post Alpbach Workshop hosted by the Space Research Institute (IWF) of the Austrian Academy of Sciences (ÖAW) in Graz; we would like to thank the Institute's Director, Wolfgang Baumjohann for hosting it. We acknowledge the financial support of ESA's Education Office that made the event possible, and the support from FFG, ISSI and Austrospace. Siegfried Eggl and Nicola Sarda acknowledge the support from the European Science Foundation (ESF) for presenting the SEARCH mission concept during the ESF Exploratory Workshop on "Observation, Characterization and Evolution of Habitable Exoplanets and their Host Stars" in Bairisch Kölldorf, Austria, Nov. 2009. Finally Siegfried Eggl would like to acknowledge the support of the Austrian FWF project P20216 and Veresa Eybl wants to acknowledge the support of the Austrian FWF project P18930-N16. Nicolas Sarda would like to acknowledge financial support from Astrium Ltd.